# Visualizing Periodic Stability in Studies:

# the moving average meta-analysis (MA$^2$)

Konstantinos Pateras[1*], Suhail A. R. Doi[2], Kit CB Roes[3], Polychronis Kostoulas[1]

[1] Department of Public and One Health, Medical school, University of Thessaly, Greece, kostas.pateras@gmail.com, pkost@uth.gr

[2] Department of Population Medicine, College of Medicine, QU Health, Qatar, University, Doha, Qatar, sdoi@qu.edu.qa

[3] Department of Health Evidence, Radboud University Medical Centre, Section Biostatistics, Nijmegen, The Netherlands, kit.roes@radboudumc.nl

*Konstantinos Pateras, Laboratory of Epidemiology, Applied Artificial Intelligence and Biostatistics, Department of Public and One Health, Medical school, University of Thessaly, Greece, kostas.pateras@gmail.com*

**Acknowledgements**

**Data availability**
Data can be found in Figures 1 and 2 and in their respective manuscripts.

**Funding**
Suhail Doi was supported by Program Grant #NPRPBSRA01–0406–210030 from the Qatar National Research Fund. The findings herein reflect the work and are solely the responsibility of the authors.

**Conflict of interest disclosure**
The authors have no conflicts of interest.

# Abstract

Objective: Stability of clinical benefits are often visually explored and formally analyzed through a cumulative meta-analysis. This approach can be enhanced further to enable identification of periodic stability of the meta-analytical result, defined as the variation of meta-analytical results across different time-periods.

Study design and settings: We describe the development of a moving average meta-analysis to aid visualization of periodic stability in meta-analysis. We illustrate the approach via two real-life examples.

Results: Moving average meta-analysis was shown to enable a dynamic view of data, by facilitating an understanding on how estimates evolve over different periods in time. There is better visualization of periodic changes in estimated heterogeneity, overall effects, influential studies and outliers all of which can allow examination of the impact of different periods towards the overall effect.

Conclusion: The moving-average meta-analysis framework provides a visual add-on to the traditional cumulative meta-analysis leading to enhanced visual exploration of periodic (in) stability.

# Introduction

Meta-analysis and other research synthesis methods aim to combine the accumulated scientific evidence from diverse individual studies that address a specific research question. The body of evidence is not static and evolves over time, due to random and systematic error (attributed to differences in research methods or participant characteristics) related deviations from the population parameter of interest[1]. As clinical studies accrue, the synthesized effect size may evolve towards a constant value (stability) or may continue to be revised over time. Consequently, any policy recommendations formed, based on this pool of evidence may benefit from assessment of the stability of a meta-analytical result. Still, there may also be a need (especially within infectious diseases) to assess periodic stability of meta-analytic estimates in research areas where data are rapidly generated[2].

The traditional method to examine meta-analysis for stability would be the cumulative meta-analysis (CMA) or cumulative sum (CUSUM) plots, which encapsulate the evidence accumulated up to the most recent time point[3–5]. These methods however need to be updated if the intent of decision-makers is to examine whether the continual accumulation of evidence is associated with periodic variation, such that the estimated effect may be different in different time periods. In this situation, a meta-analysis should ideally be conducted over several shorter time-frames, thereby excluding older or newer and future studies. In this second context, we propose an index of periodic stability of studies within different periods - the Moving Average Meta-Analysis (MA$^2$ or MAMA) approach, featuring a flexible-sized running window dependent on the specific research question. To introduce MA$^2$ as an extension of the traditional CMA we deploy two meta-analyses as motivating examples, and discuss the choice of window size.

## Materials and methods

### Introducing the moving average meta-analysis (MA$^2$)

Temporal trends are frequently analyzed using Cumulative Meta-Analysis (CMA), as originally proposed by Lau et al.[4] The CMA process involves a step-by-step meta-analysis of studies selected through a systematic review, performed in chronological order. Each subsequent study is incorporated into the analysis, and an updated overall treatment effect is generated. To track the evolution of a treatment effect over time, studies are arranged by publication year/month, and the method computes the overall effect size alongside a 95% confidence interval at time point (t) and study (k). This computation contains a series of meta-analyses using an appropriate model[6,7], starting with the synthesis of the first two of these studies, then the first three of them, continuing in this manner until all available studies have been included in the analysis. A forest plot can be utilized to present the accumulated temporal clinical evidence, to detect changes in effect size and precision based on publication date, and to identify time points where effect estimates have stabilized. One advantage of CMA is that the inclusion of new studies enhances the sample size and depending on the assumed model, it improves the precision of the estimated effect size, resulting in narrower 95% confidence intervals and potentially robust evidence supporting one hypothesis over another. CMA thus facilitates the early establishment of a benefit-risk profile for therapeutic and preventive interventions.

Alternatively, if the intent is to judge the impact of new evidence generated at different time-points, then we need a modified approach. Stability of effects can alternatively be explored using a moving average meta-analysis (MA$^2$), as proposed herein. Similar to CMA, MA$^2$ involves a step-by-step meta-analysis of studies selected through a systematic review, performed in chronological order. Each subsequent study is included in the analysis, and an updated overall treatment effect is calculated. Yet, MA$^2$ diverges from CMA as in each step MA$^2$ includes only a fixed number of the most recent studies denoted by the "window - *l*". The term "window" refers to the number of the most recent studies pooled at time point (t). For instance, if the window is set to *l=5*, then only the latest five studies up to study (k) will contribute to the overall treatment effect at time point (t) in a meta-analysis, similarly to the moving-average concept[8,9]. To track stability and temporal changes of a treatment effect, studies are once again organized by publication year. The overall effect size, along with a 95% confidence interval, is computed up to the earliest study within the set window (l), excluding all studies older than this range. An appropriate meta-analysis model can then be performed by combining the first two of these studies, then the first three of them, and so on, until the number of pooled studies reaches the window size. After this point, per each newer study included, an older study is excluded from the analysis. In this way, we can estimate the periodic variation in meta-analytical results. A forest plot can be used to summarize temporal evidence, much like in CMA analysis. A potential downside of MA$^2$ is the exclusion of older studies, which restricts the overall sample size and influences the precision of the estimated overall effect size. This can result in narrower or wider 95% confidence intervals than those produced by a CMA. Conversely, a positive aspect of MA$^2$ is that, as more evidence accumulates, the influence of newer studies on the results is amplified compared to CMA.

**Choice of meta-analysis model and method of computation**

In this study, we will utilize the IVhet estimator for meta-analysis[10], that has been shown to perform robustly as it avoids unnecessary normality approximations, as compared to the common random-effects model [6,7,11], whereby, the variance of a CMA when it includes study (k), is influenced by the following equation:

$$\widehat{Var}(\hat{\delta}_{CMA\,(k)}) = \sum_{i=1}^{k}\left[\left(\frac{\frac{1}{\hat{\sigma}_i^2}}{\sum_{i=1}^{k}\frac{1}{\hat{\sigma}_i^2}}\right)^2 (\hat{\sigma}_i^2 + \hat{\tau}_{1,\,k}^2)\right] \quad [1]$$

Where $\hat{\tau}_k^2$ is the between studies variance by the method of moments only and the study weights are given by $\hat{w}_i = (\hat{\sigma}_i^2)^{-1}$ denoting inverse-variance weights. $\hat{\delta}_{CMA\,(k)} = \left[\sum_{i=1}^{k}\hat{w}_i\hat{\delta}_i\right]/\left[\sum_{i=1}^{k}\hat{w}_i\right]$, where $\hat{\delta}_i$ denotes an estimate of the study-specific effect. Depending on the type of the measure of interest (i.e. proportion, mean difference etc) the estimated study-specific variance is estimated based on a specific equation all of which are a function of the study-specific sample sizes[12].

The variance of a MA[2,] after including study (k), is given by the following expression

$$\widehat{Var}(\hat{\delta}_{MA^2\,(k,\,l)}) = \sum_{i=k-l}^{k}\left[\left(\frac{\frac{1}{\hat{\sigma}_i^2}}{\sum_{i=k-l}^{k}\frac{1}{\hat{\sigma}_i^2}}\right)^2 (\hat{\sigma}_i^2 + \hat{\tau}_{k-l,k}^2)\right], 1 \leq l < k \quad [2]$$

,where *l is the moving average window size*. In MA$^2$, compared to CMA, the number of studies included in the analysis, after the first *l* studies, remains constant and equal to *l* and the precision (not accounting for τ) should remain relatively similar between time points for the MA$^2$, therefore, $\hat{\delta}_{MA^2\,(k)} = \left[\sum_{i=k-l}^{k}\hat{w}_i\hat{\delta}_i\right]/\left[\sum_{i=k-l}^{k}\hat{w}_i\right]$. In MA$^2$, τ is estimated by *l* number of studies. By utilizing the IVhet model that avoids normality assumptions, we need not worry about the window being, if needed, relatively small for the inference on the heterogeneity parameter, as would be the case with a random effects meta-analysis[13]. If heterogeneity is of interest, a reasonable value for the MA$^2$ window should be chosen, when possible, as MA of very few studies would not provide meaningful and unbiased results[14,15]. As expected, as the window (l) becomes larger, the MA$^2$ method converges to a CMA, while when *l=1*, then the equation reduces to the latest available study (k) *reproducing the* study-specific forest plot.

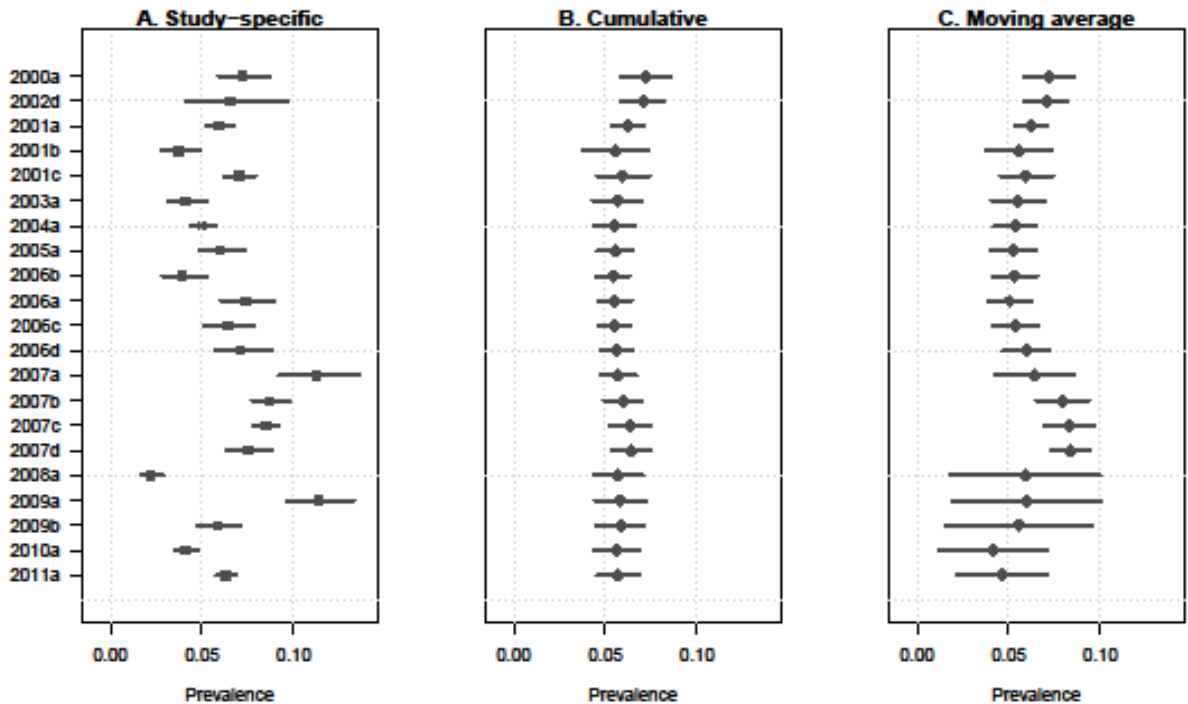

Figure 1. Forest plots presenting A) the study-specific effects reported from Willcut 2012[16] and B) CMA and C) $\mathrm{MA}^2$ ($l=5$) overall effects, each line depicts either a study (1A), a step of the CMA analysis (1B) or a step of the $\mathrm{MA}^2$ analysis (1C).

## Results - Motivating examples

### Willcut's - Attention deficit/hyperactivity disorder (ADHD) data

The first example focuses on a comprehensive meta-analysis to estimate the prevalence of attention-deficit/hyperactivity disorder (ADHD) diagnosed based on the Diagnostic and Statistical Manual of Mental Disorders, fourth edition (DSM-IV)[16]. A main challenge in estimating the overall prevalence of ADHD stems from the fact that its symptoms frequently overlap with those of other psychiatric disorders. The authors undertook a systematic literature review, including a total of 86 studies involving children and adolescents (total sample size of 163,688) that met the inclusion criteria for a meta-analysis[16], the analysis was updated in 2002 and 2012[16].

Position of Table 1

In Figure 1, a re-analysis of 21 studies from Willcutt's work is presented. The studies chosen for re-analysis specifically focused on full DSM-IV ADHD and reported prevalences based on questionnaires completed by parents. The presented re-analysis synthesizes logit transformed prevalence proportions.

The $\mathrm{MA}^2$-related approach clearly demonstrates a lack of periodic stability in the effect estimates, which the CMA does not demonstrate. Even though, the increase in heterogeneity can also be observed in the study-specific forest plots, unstable overall effects are challenging to identify when examining such study-specific plots (Figures 1A and 2A). In $\mathrm{MA}^2$ the inclusion of each subsequent study and exclusion of previous studies allows each step to be more susceptible to change, in comparison to the CMA analysis (Figures 1 and 2). This is depicted in Figure 1C, when the analysis reaches more recent study estimates and the overall estimated prevalence becomes unstable both regarding its mean and its variance.

Table 1 presents the comparison of two parameterizations of $\mathrm{MA}^2$ under two window sizes ($l_1$=5 and $l_2$=10). Influential studies' impact becomes more evident under a lower window size (l). When Waschbusch et al 2007 relatively high prevalence study is included in the analysis and when Dopfner et al 2008 lowest prevalence study is included in the analysis combined with subsequent studies, $\mathrm{MA}^2$ analysis point towards no periodic stability. As expected, as the window size increases, the behaviour of $\mathrm{MA}^2$ converges to CMA.

**Batary's et al - biology evolution data**

The efficacy of agri-environmental management in promoting biodiversity conservation on farmland has produced mixed results, leading to doubts about its overall effectiveness. In this second example, we focus on the work of Batáry et al. (2011) that examined the role of agri-environmental management schemes in conservation and environmental management[17]. A specific example was further explored in an elaborate methodological article on CMA by Kulinskaya and Mah[5]. The dataset analysed therein included species richness data extracted from 39 studies published between 1992 and 2010. The effect measure used was the standardized mean difference (SMD), where positive values indicate higher species richness in extensive (organic) fields as compared to intensive (conventional) fields. The authors focused on specific sub-studies and utilized raw data to conduct their re-analysis[5]. In their analysis, they report that an initial assessment of the forest plot reveals that the effects in the first 18 studies tend to hover slightly above zero, with Study 19 (S 20062) and Study 33 (2002a/2) standing out as outliers.

In Figure 2, a re-analysis of 39 studies from Batary's work as summarized in Kulinaska and Mah is presented[5,17]. The forest plots are presented containing the CMA and $\mathrm{MA}^2$ estimates. $\mathrm{MA}^2$ results in wider or narrower 95% confidence intervals than CMA, that depends on the studies contained in the current meta-analysed set of *l* studies. In this example there is evidence of periodic stability due to precise studies entering in certain periods, which the CMA clearly does not portray.

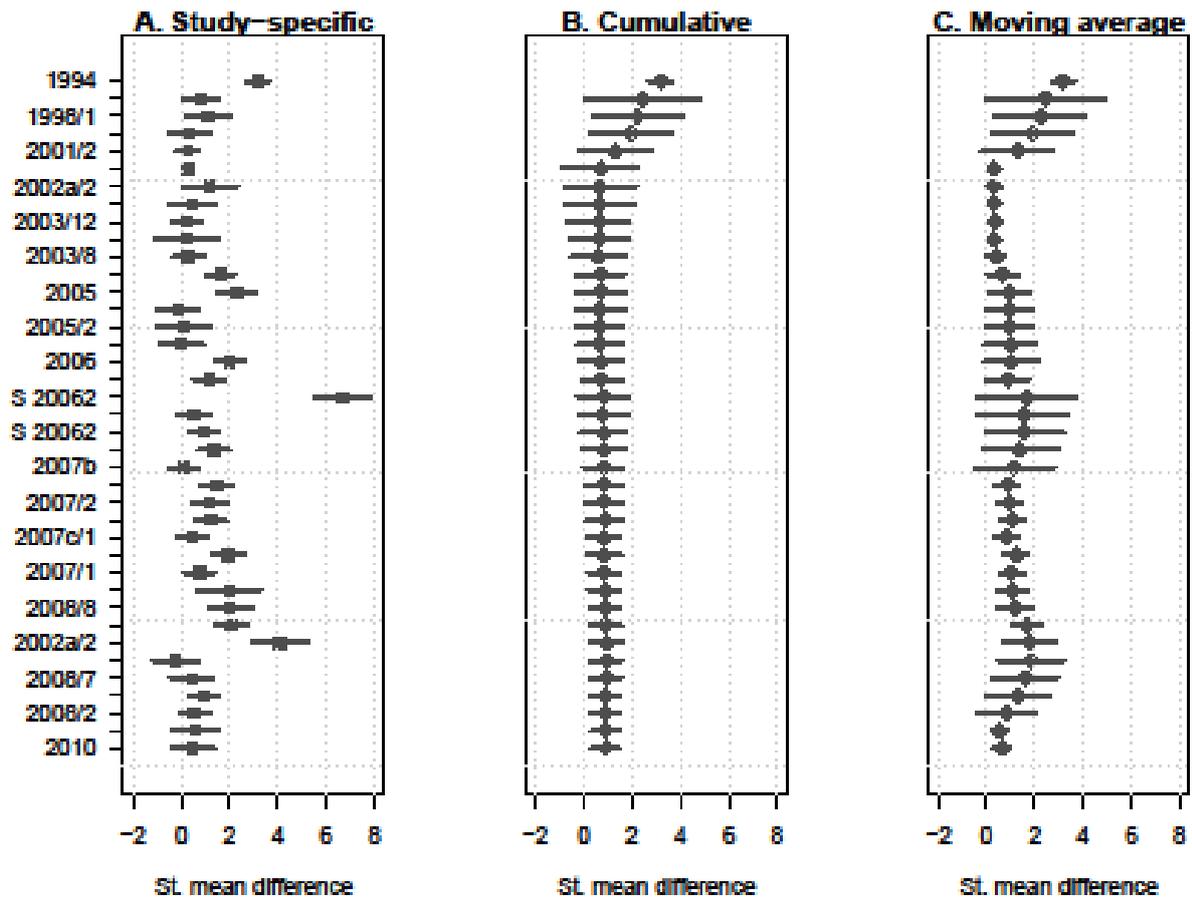

Figure 2. Forest plots presenting A) the study-specific from Kulinskaya and Mah 2022[5], B) CMA and C) $MA^2$ ($l$=5) overall effects, each line depicts either a study (1A), a step of the CMA analysis (1B) or a step of the $MA^2$ analysis (1C).

## Discussion

This manuscript introduces the concept of a moving-average meta-analysis. Up until now, the examination of periodic stability in a meta-analysis was not discernable through the application of cumulative meta-analysis (CMA). The moving average is a well-established notion across multiple scientific fields, thus, the core mechanism to perform our proposed analysis is very robust[8,9]. A $MA^2$ would be helpful for specific cases, where periodic stability needs to be clearly visualized, examining the most recent $l$ studies up to time-point $t$. $MA^2$ does not focus on establishing adequate evidence to reject a null hypothesis, but instead the focus is on identifying periods that differ in the evolution of an effect, thus, a $MA^2$ analysis is not required to be prospectively planned. This means that if the analysis pools $k$ studies sequentially, we would like to plot the pooled effects over each window, and it does not matter what the type I error is, which therefore translates to a different approach than sequential (meta) analysis of randomized clinical trials[18,19].

A $MA^2$ has some useful traits as an extension to CMA; for instance, in the case of high heterogeneity, also related to temporal changes, the heterogeneity of CMA-related methods may be increased, even if newer studies produce similar effect estimates. In the proposed approach, the IVhet model CI will allow for overdispersion related to heterogeneity of studies in the window selected. Thus the time trend will be able to also depict the extent of heterogeneity within each window and how it changes.

The MA² approach may share some equivalence with the use of a meta-regression and time as an exploratory variable; nonetheless, the concept of introducing a specific window and exploring its impact is unique to the proposed approach. This trait of MA² comes with added responsibility, as users of MA² should be careful selecting a proper-sized window (*l*). For instance, a MA² may be challenging when the window size is small due to the mis-estimation of heterogeneity and other related issues stemming from the meta-analysis of small populations literature[20,21], or less informative when (*l*) is very close to (*k*). Similarly to CMA, MA² can aid researchers to early identify specific biases that appear early in a series of studies, such as the novelty bias or the early-publication bias. Evidently, MA² shares some equivalence with smoothing techniques commonly applied in the field of quality control and time-series analysis[9]. We decided not to focus on such non-uniform weight functions, as they would produce a type of standardization analysis, rather than a meta-analysis[22].

The novelty in the MA² approach is the ability to address periodic stability of estimates which may be useful in clinical decision making, as it can offer concurrent insights regarding a consistent or a fluctuating effect. MA² (in) stability evaluation can 1) result in timely adjustment of patient management strategies and 2) suggest timely reallocation of clinical resources. In addition, an MA² can provide new insights to specific areas of interest for meta-analysis, such as the small-study effects and publication bias, through dynamically exploring the periodic course of evidence through time. Indeed, even conventional subgroup analysis under MA² acquires a different meaning, that of the (in) stability across subgroups and their clinical implications. MA² can also provide further insights in areas where conventional meta-analysis technically struggles, such as in very high heterogeneity exploration, informative missing outcomes and rare events that lead to zeros and biases. By focusing formally on specific windows, a MA² can guide practitioners on whether periodic differences exist and if they are of interest or not. Still, interpretation of (in) stability should be made in accordance with clinical practice.

## What is new?
Key findings

- A novel approach is proposed on modelling meta-analytical data via the use of a moving average (MA²), this approach is impacted by a specific number of historical studies at a given time point.

What this adds to what is known related to methods research within the field of clinical epidemiology

- Methods such as a cumulative meta-analysis are commonly used to assess and track the evolving body of evidence over time, but it is difficult to use the latter to make claims regarding periodic stability of effect estimates.
- An MA² makes possible the visual evaluation of periodic stability of effect estimates over time.

What is the implication, what should change now

- The proposed method provides insights in areas where conventional meta-analyses and/or standard forest plots will struggle, such as, influential studies, informative missing outcomes, and rare events by demonstrating their impact over different periods of analysis.

|  | $MA^2_5$ | | $MA^2_{10}$ | |
|---|---|---|---|---|
| Author - Year | Prevalence | 95% CI | Prevalence | 95% CI |
| Willcutt 2011 | 0.046 | (0.07,0.02) | 0.058 | (0.08,0.04) |
| Anselmi 2010 | 0.042 | (0.07,0.01) | 0.057 | (0.08,0.03) |
| Cho 2009 | 0.056 | (0.10,0.02) | 0.062 | (0.09,0.04) |
| Tuvblad 2009 | 0.060 | (0.10,0.02) | 0.061 | (0.09,0.03) |
| Dopfner 2008 | 0.059 | (0.10,0.02) | 0.059 | (0.08,0.03) |
| Bauermeister 2007 | 0.084 | (0.09,0.07) | 0.069 | (0.08,0.05) |
| Smalley 2007 | 0.083 | (0.10,0.07) | 0.066 | (0.08,0.05) |
| Froehlich 2007 | 0.079 | (0.09,0.06) | 0.061 | (0.07,0.05) |
| Waschbusch 2007 | 0.064 | (0.09,0.04) | 0.056 | (0.07,0.04) |
| Bird1 2006 | 0.060 | (0.07,0.05) | 0.055 | (0.06,0.05) |
| Bird2 2006 | 0.054 | (0.07,0.04) | 0.055 | (0.06,0.05) |
| Skounti 2006 | 0.050 | (0.06,0.04) | 0.055 | (0.06,0.05) |
| Egger 2006 | 0.053 | (0.07,0.04) | 0.054 | (0.06,0.04) |
| Neuman 2005 | 0.052 | (0.06,0.04) | 0.055 | (0.06,0.05) |
| Rasmussen 2004 | 0.053 | (0.06,0.04) | 0.055 | (0.07,0.04) |
| Costello 2003 | 0.055 | (0.07,0.04) | 0.057 | (0.07,0.04) |
| Graetz 2001 | 0.060 | (0.07,0.05) | 0.060 | (0.07,0.05) |
| Kroes 2001 | 0.055 | (0.07,0.04) | 0.055 | (0.07,0.04) |
| Todd 2001 | 0.062 | (0.07,0.05) | 0.062 | (0.07,0.05) |
| Benjasuwantep 2002 | 0.070 | (0.08,0.06) | 0.070 | (0.08,0.06) |
| Willoughby 2000 | 0.072 | (0.09,0.06) | 0.072 | (0.09,0.06) |

Table 1. Pooled prevalences and confidence intervals of the moving average meta-analysis with window equal to $l_1=5$ and $l_2=10$ utilizing prevalence data from Willcut 2012[16]. The exploration of windows equal to 5 and 10 is set to compare two distinct sizes of windows. The actual selected window for performing a $MA^2$ should take into account the number of studies and the level of sensitivity that the researcher would like to introduce in the $MA^2$ analysis.